%% file: vprec_libm_short_hal.tex
\documentclass[conference]{IEEEtran}
\IEEEoverridecommandlockouts
\usepackage{cite}
\usepackage{amsmath,amssymb,amsfonts}
\usepackage{algorithmic}
\usepackage{graphicx}
\usepackage{textcomp}
\usepackage{xcolor}

\usepackage{tikz}
\usepackage{pgfplots}
\pgfplotsset{compat=newest}

\usetikzlibrary{calc, shapes, shapes.geometric, arrows, positioning}

\usepackage{comment}

\usepackage{caption}
\usepackage{subcaption}

\usepackage{adjustbox}
\usepackage{float}

\usepackage{hyperref}

\hypersetup{
	colorlinks=true,
	linkcolor=blue,
	filecolor=magenta,      
	urlcolor=cyan,
	pdftitle={Custom-Precision Mathematical Library Explorations for Code Profiling and Optimization},
	bookmarks=true,
}

\usepgfplotslibrary{statistics}

\pgfdeclarelayer{edgelayer}
\pgfdeclarelayer{nodelayer}
\pgfsetlayers{edgelayer,nodelayer,main}

\tikzstyle{line} = [draw, -latex', very thick]
\tikzstyle{dottedLine} = [draw, -latex', semithick, dotted]
\tikzstyle{dottedLineSimple} = [draw, -, semithick, dotted]

\tikzstyle{dataStructure} = [rectangle, draw, fill=gray!10, text centered, minimum height=4em, thick]
\tikzstyle{process} = [rectangle, draw, rounded corners, fill=gray!5, text centered, node distance=3em, minimum height=2em, thick]

\makeatletter
\pgfdeclareshape{document}{
		\inheritsavedanchors[from=rectangle]
		\inheritanchorborder[from=rectangle]
		\inheritanchor[from=rectangle]{center}
		\inheritanchor[from=rectangle]{north}
		\inheritanchor[from=rectangle]{south}
		\inheritanchor[from=rectangle]{west}
		\inheritanchor[from=rectangle]{east}
		\backgroundpath{
				\southwest \pgf@xa=\pgf@x \pgf@ya=\pgf@y
				\northeast \pgf@xb=\pgf@x \pgf@yb=\pgf@y
				\pgf@xc=\pgf@xb \advance\pgf@xc by-10pt
				\pgf@yc=\pgf@yb \advance\pgf@yc by-10pt
				\pgfpathmoveto{\pgfpoint{\pgf@xa}{\pgf@ya}}
				\pgfpathlineto{\pgfpoint{\pgf@xa}{\pgf@yb}}
				\pgfpathlineto{\pgfpoint{\pgf@xc}{\pgf@yb}}
				\pgfpathlineto{\pgfpoint{\pgf@xb}{\pgf@yc}}
				\pgfpathlineto{\pgfpoint{\pgf@xb}{\pgf@ya}}
				\pgfpathclose
				\pgfpathmoveto{\pgfpoint{\pgf@xc}{\pgf@yb}}
				\pgfpathlineto{\pgfpoint{\pgf@xc}{\pgf@yc}}
				\pgfpathlineto{\pgfpoint{\pgf@xb}{\pgf@yc}}
				\pgfpathlineto{\pgfpoint{\pgf@xc}{\pgf@yc}}
			}
	}
\makeatother

\tikzstyle{doc}=[%
	draw,
	thick,
	align=center,
	color=black,
	shape=document,
	minimum width=20mm,
	minimum height=28.2mm,
	shape=document,
	inner sep=2ex,
]

\def\BibTeX{{\rm B\kern-.05em{\sc i\kern-.025em b}\kern-.08em
    T\kern-.1667em\lower.7ex\hbox{E}\kern-.125emX}}
\begin{document}
\sloppy

\title{
	Custom-Precision Mathematical Library Explorations for Code Profiling and Optimization
}

\author{\IEEEauthorblockN{David Defour\IEEEauthorrefmark{4}, Pablo de Oliveira Castro\IEEEauthorrefmark{1}\IEEEauthorrefmark{2}, Matei I\c{s}toan\IEEEauthorrefmark{1}\IEEEauthorrefmark{2}, and Eric Petit\IEEEauthorrefmark{2}\IEEEauthorrefmark{3}}
\IEEEauthorblockA{\IEEEauthorrefmark{1} University of Versailles -- Li-PaRAD,
Email: \{pablo.oliveira, matei.istoan\}@uvsq.fr}%
\IEEEauthorblockA{\IEEEauthorrefmark{2} Exascale Computing Research, ECR}%
\IEEEauthorblockA{\IEEEauthorrefmark{3} Intel Corporation,
Email: eric.petit@intel.com}%
\IEEEauthorblockA{\IEEEauthorrefmark{4} University of Perpignan,
Email: david.defour@upvd.fr}}

\maketitle

\begin{abstract}
The typical processors used for scientific computing have fixed-width data-paths. 
This implies that mathematical libraries were specifically developed to target each of these fixed precisions (binary16, binary32, binary64).
However, to address the increasing energy consumption and throughput requirements of scientific applications, library and hardware designers are moving beyond this one-size-fits-all approach. 
In this article we propose to study the effects and benefits of using user-defined floating-point formats and target accuracies in calculations involving mathematical functions. 
Our tool collects input-data profiles and iteratively explores lower precisions for each call-site of a mathematical function in user applications. 
This profiling data will be a valuable asset for specializing and fine-tuning mathematical function implementations for a given application. 
We demonstrate the tool's capabilities on SGP4, a satellite tracking application. 
The profile data shows the potential for specialization and provides insight into answering where it is useful to provide variable-precision designs for elementary function evaluation.

\end{abstract}


\section{Introduction}\label{sec:intro}
A growing interest to adapt floating-point formats to the real needs of applications is becoming ever more ubiquitous.
This process has been successfully conducted by the AI community which has settled on the BF16~\cite{intel-bfloat16} and fp16~\cite{nvidia-float16} formats, in order to increase performance and efficiency.
Similar benefits have been achieved in other domains~\cite{Chatelain2019automatic, haidar2018design} by reducing the precision of basic operations ($+,-,*,/$) and harnessing hardware-support for multiple internal floating-point formats.

Oberman et al.~\cite{565590} demonstrate that neglecting optimizations of infrequent operations, such as division and square root, can severely impact performance.  
We believe that elementary functions should not be neglected either when optimizing for mixed-precision.
Even though elementary functions are not widely available in hardware and infrequently used in applications, their impact can be important.
HPC Patmos neutronic solver~\cite{brun2017patmos} spends $70\%$ of the execution time in the mathematical library (\textit{libm}) functions. 
This confirms similar observations by CERN~\cite{piparo:hal-01116217} on their HPC codes.

Various mathematical libraries specialize evaluation schemes for different accuracy/performance trade-offs~\cite{mkl}. 
Recent developments such as {\it metalibm}~\cite{7203797} automatically generate elementary functions to best fit the hardware and accuracy constraints.
These works highlight that a trade-off can be explored between performance, accuracy and precision.

In this paper, we propose a tool for collecting input intervals and output required precision profiles from real applications in order to guide the design of specialized mathematical libraries.  
Indeed, considering limited input data ranges and application-focused output accuracy could drastically influence the implementation performance. 
We demonstrate the tool's capabilities on SGP4, a satellite tracking application. 
The profile data shows the potential for specialization and provides insight into where it is useful to provide variable-precision designs for elementary function evaluation.

\section{Related Works}\label{sec:related_work}
Contrary to basic operations, properties of elementary function are not standardized mainly because the correctly rounded property is difficult to achieve~\cite{10.1145/2701.356101,Defour:2004sl}.
As a consequence, there are numerous available implementations of such functions, either in software or hardware, each representing a different trade-off between, accuracy, performance, hardware requirements and programming language. 
The most notable mathematical library embedding different trade-offs are the Vector Mathematical Functions from Intel's MKL library~\cite{mkl}, which offers three accuracy modes: High-/Low-accuracy and Enhanced Performance. 
Another example are Nvidia's GPUs, which embed dedicated hardware for fast approximation of some functions and software implementation of more accurate and larger input range versions~\cite{4459293}. 
This has led to the OpenCL 2.2 standard which defines the requirements in terms of accuracy of mathematical functions from {\tt half}  to {\tt double}~\cite{Munshi:2016:OCS}.

Developing and maintaining multiple implementations for each function is a daunting endeavor.
Several tools have been proposed to automate this task either, for hardware or software implementation of such functions~\cite{oai:prunel.ccsd.cnrs.fr:ensl-00506125,7203797}.

Porting the concept of memoization to mathematical functions has been explored in~\cite{collange:hal-00202906,suresh:tel-01410539} where the authors investigated how considering real input-data profile can be used to optimize the evaluation. 
However, they did not evaluate the potential decrease in accuracy.

\section{Simulating Variable Precision and Range for Mathematical Functions}\label{sec:vprec}

Our approach to simulate mathematical functions with reduced range/precision is twofold: first we transparently interpose calls to mathematical functions, then the \texttt{VPREC-libm} library computes the result in a reduced precision.

\subsection{Library Call Interposition}\label{sec:vprec:library}

In Linux, the dynamic loader offers the possibility to intercept dynamic
library calls, so that a custom library is called instead.  This is achieved by
setting the \texttt{LD\_PRELOAD} environment variable. This interception method works out-of-the-box with a compiled binary and is transparent to the user. 

We use it to replace
standard calls to the \texttt{libm} with custom calls to our own
\texttt{VPREC-libm} library which simulates non-standard precisions and range. This approach is flexible, but has two limitations:
\begin{itemize}
  \item it is not applicable to statically linked programs: for those, the user must manually re-link the program against the \texttt{VPREC-libm};
  \item it only intercepts library calls; to ensure that we intercept all operations we disable compiler optimizations which replace calls by hardware instrinsics (such as \texttt{sqrtsd} assembly instruction in IA-64). Fortunately, the \texttt{-fno-builtin} flag disables these optimizations in most standard compilers.
\end{itemize}

\input{vprec_libm_overview.tex}
\vspace{-10mm}
\subsection{Implementation of the VPREC-libm}\label{sec:vprec:implementation}

The interposed mathematical call is handled by \texttt{VPREC-libm}, which returns the result in the target floating-point format.  
For example in double precision, one can customize the bit length of the pseudo-exponent $r \in [1,11]$ and the pseudo-mantissa $p \in [0,52]$. 

VPREC-libm operates in two steps. First, it computes a \texttt{binary128}
result $\widetilde{z}$  by calling the corresponding mathematical function from the GCC's \texttt{libquadmath}.

Then, $\widetilde{z}$ is converted to the target format using
Verificarlo-VPREC~\cite{Chatelain2019automatic}.
If $\widetilde{z}$ is representable in the
target range, a faithfully rounded result at target precision $p$ is returned.
If $\widetilde{z}$ is outside the target range, VPREC returns $\pm \infty$ for overflows and $\pm 0$ for underflows. Rounding is achieved by adding a ulp at precision $p+1$ followed by a truncation ($\lfloor \widetilde{z} + 2^{e_z-p-1} \rfloor_p$).

\subsection{Exploring Precision Requirements Using VPREC-libm}\label{sec:optimization}

\texttt{VPREC-libm} can be used in two modes: \textit{profiling}, and \textit{execution}.
In profiling mode, \texttt{VPREC-libm} creates a profile of the executed code.
For each call, it updates the boundaries of the operands and output intervals, the number of occurrences for the unique program address and stack trace from which the call is made.
This information is aggregated and processed to produce execution statistics.

In execution mode, \texttt{VPREC-libm} accepts a configuration file which specifies the floating-point formats to use at each call-site.
An initial config file is generated automatically after a profiling run, containing information for all encountered call-sites, which the user can later modify as it is required.

Taking advantage of this functionality, we developed a method for exploring the optimization potential of a given floating-point code.
On a broad scale, we perform a dichotomic search for the minimal output precision of the \texttt{VPREC-libm} functions which meets the user-specified correctness criteria.
This search is applied sequentially per call-site and converges in logarithmic time, in the size of the mantissa.
As we are dealing with a vast search-space, the order in which this optimization is performed is quite important. According to Amdahl's law, optimizing the code in which most of the time is spent results in the biggest performance gains.
Therefore we prioritize the call-site exploration by call frequency as a heuristic.
Figure \ref{fig:vprec-libm_overview} summarizes this process.

\section{Experimental Results}\label{sec:exp_results}

The example discussed throughout this section illustrates the potential of our proposed method, applied to a real-world astrophysics application used to predict the position and the velocity of Earth-orbiting objects (most notably satellites).

\subsection{Satellite Tracker: SGP4}\label{sec:exp_results:spg4}

In order to track the position and the velocity of satellites, a common technique is to use \textit{simplified general perturbations (SGP)} models, which predict the influence of drag, the Earth's shape, as well as that of the sun and the moon on the trajectory.
Work on SGP started in the 1960's, but at the beginning of the 1980s NORAD\footnote{North American Space Defense Command} released the equations and the source code to predict satellite positions~\cite{Hoots80}. 
NORAD maintains and periodically refines data sets on all resident space objects, ensuring the accuracy of trajectory predictions.
The data-sets were made available to users, through NASA, being the only source of readily available orbital data.

However, a user could not just go ahead and use any prediction model she wished, she had to use the same model employed to generate the data-set, even if the user's choice might have had better performance, an aspect highlighted in~\cite{Hoots80}.
This made SGP, especially its SGP4/SDP4 variants, commonplace among users wishing to use NORAD's orbital data, distributed as \textit{two-line element (TLE)} sets.
\textit{Near-Earth} objects (period less than $225$ minutes) are tracked using SGP4, while \textit{deep-space} objects (period over $225$ minutes) are tracked using SDP4, which also models the gravitational effects of the moon and the sun and certain Earth harmonics.

In the period following the original release of SGP, a multitude of code variations came to exist, making interoperability and compatibility an issue. For experimental purposes, we will use the SGP4 version documented in~\cite{Vallado2006}, which is a community effort to keep a version up to date with the models and TLE data-sets used by NORAD.
The mathematical models used throughout are those presented in~\cite{Hoots2004}.
The \texttt{C++} version used for the experiments is the one provided by \href{https://www.celestrak.com}{CelesTrak}, available \href{https://celestrak.com/publications/AIAA/2006-6753/}{online}\footnote{httsp://celestrak.com/publications/AIAA/2006-6753/}.
Only minimal changes were made, in order to ensure that the program runs correctly on Linux.
This has not affected the program outputs, as verified against the provided test-suites.

\subsection{Results}\label{sec:exp_results:results}

We applied our method to the data-set \texttt{sgp4-all.tle} discussed in~\cite{Vallado2006}.
Figure~\ref{fig:all_sat} illustrates our profiling and precision exploration results for the $50$ {\it libm} call-sites with the most calls.
At the top we show the number of occurrences for different  {\it libm} call-sites.
In the middle graph we show for each call-site the dynamic range of the input data, as used in the original code, extracted from the profile produced by \texttt{VPREC-libm}.
Finally, in the bottom graph we show for each call-site the required precision of the outputs determined by our exploration with the \texttt{VPREC-libm} method.

We can observe that the second and third call-sites are almost twice more frequent than the next ten entries.
The output precision cannot be significantly decreased for the two first call-sites, as shown in the bottom sub-figure. 
Indeed they occur at the very end of the algorithm, directly influencing the outputs.
On the other hand, the dynamic range of the input is quite reduced at these two call-sites. 

Analyzing the code, we notice that these calls to \texttt{sin()} and \texttt{cos()} take the same argument, as they are part of a rotation.
Therefore, they could be replaced with a call to \texttt{sincos()}, effectively reducing their combined workload by almost a factor of two, similar to what is done in~\cite{Markstein2003}.

Analyzing the rest of the call-sites, we can observe two general trends.
The first are call-sites where the required precision is close to the default one; here we only manage to save $5-6$-bits.
The second are call-sites where the required precision hovers around the $28$-bits mark.
A plausible explanation for these results can be found through a bit of computer-science archaeology.
SGP4 was first developed in \href{https://en.wikipedia.org/wiki/Fortran\#FORTRAN\_IV}{FortranIV}\footnote{https://en.wikipedia.org/wiki/Fortran\#FORTRAN\_IV}, on a \href{https://en.wikipedia.org/wiki/Honeywell\_6000\_series}{Honeywell-6000}\footnote{https://en.wikipedia.org/wiki/Honeywell\_6000\_series} series computer~\cite{Hoots80}.
This machine had $36$-bit words, so floating-point numbers had a $8$-bit exponent and a $28$- or $64$-bit significand, for single- or double-precision, respectively.
As noted in~\cite{Vallado2006}, the code originally used a mix of single- and double-precision computations.
With the evolution of the underlying hardware, the code was moved to double-precision throughout, which made for a smoother behavior, but did not improve the accuracy.
These observations are indeed coherent with our findings on optimizing the precision of the outputs of the mathematical functions.

Figure \ref{fig:5_sat} shows the results of our method on to a data-set containing just the satelite number $5$, which is the first satellite in the \texttt{sgp4-all.tle} dataset.
This is a near-Earth satellite, which means that its trajectory is tracked using SGP4, not SDP4.
Its period, perigee and eccentricity ensure that no corner case is triggered in the model.
The only particularity of this example is that it uses the \textit{TEME}\footnote{True Equator, Mean Equinox} coordinate system, which requires a conversion to a more standard coordinate systems.
Indeed, avoiding exceptional cases in the model shows that considerably lower precisions can be used throughout.
The trends for the precision of the math functions' output remains mostly the same.
We notice, that over a third of the functions could be evaluated in single-precision, requiring at most $23$-bits of precision at the output.

It should be noted that the x-axis indices in Fig. \ref{fig:all_sat} and \ref{fig:5_sat} do not necessarily match, as the execution paths can differ, due to the different nature of the two data-sets.

\begin{figure*}

\begin{subfigure}{1.0\textwidth}
\tiny
\centering

\scalebox{0.7}{
	\hspace*{-1.7em}%
	\input{all_sat_freq.tex}
	\label{fig:all_sat_freq}
}
\end{subfigure}%
\vspace*{-0.15em}%

\hspace*{-0.35em}%
\begin{subfigure}{1.0\textwidth}
\tiny
\centering
\hspace*{-1.5em}%
\scalebox{0.7}{
	\input{all_sat_ranges2.tex}
	\label{fig:all_sat_range}
}
\end{subfigure}%
\vspace*{-0.15em}%

\hspace*{-0.35em}%
\begin{subfigure}{1.0\textwidth}
\tiny
\centering	
\scalebox{0.7}{
	\input{all_sat_prec.tex}
	\label{fig:all_sat_preq}
}
\end{subfigure}%

\vspace*{-0.75em}%
\caption{Analysis of calls to libm for all satellites data-set}
\label{fig:all_sat}

\end{figure*}

\begin{figure*}
	
\begin{subfigure}{1.0\textwidth}
\tiny
\centering
\scalebox{0.7}{
	\input{5_sat_freq.tex}
	\label{fig:5_sat_freq}
}
\end{subfigure}%
\vspace*{-0.15em}%

\hspace*{-0.5em}%
\begin{subfigure}{1.0\textwidth}
\tiny
\centering
\scalebox{0.7}{
	\input{5_sat_ranges2.tex}
	\label{fig:5_sat_preq}
}
\end{subfigure}%
\vspace*{-0.15em}%

\hspace*{-0.35em}%
\begin{subfigure}{1.0\textwidth}
\tiny
\centering
\scalebox{0.7}{
	\input{5_sat_prec.tex}
	\label{fig:5_sat_range}
}
\end{subfigure}%

\vspace*{-0.75em}%
\caption{Analysis of calls to libm for satellite 5 data-set}
\label{fig:5_sat}	
	
\end{figure*}

\section{Conclusion and Future Work}\label{sec:future_work}\label{sec:future_work:concepts}\label{sec:future_work:pgo}

In this paper, we focus on providing a software tool and methodology to profile the mathematical library usage in a full scale application. 
The objective is to measure the potential and drive future {\it ad-hoc} optimizations of the math library.

Usually, elementary functions are implemented following a four step scheme~\cite{7203796}: special-case handling, argument reduction, reduced domain splitting and interpolation (\textit{e.g.} polynomial or iterative).

When limiting the input domain, the first two steps can be optimized. 
Furthermore, reducing the required accuracy and input domain may lower the interpolation complexity.
It can also diminish the implementation cost in special purpose architecture designs or re-programmable architectures (FPGA).

Rewriting {\it ad-hoc} custom elementary functions with a target accuracy on a given input interval is a costly and error prone task. 
We propose to explore existing tools to assist or automate these optimizations and measure emprical speedup on real use cases.
An easy first step would be to automatically select the best fitting implementation among existing libraries, such as Intel MKL VML~\cite{mkl}. 
Finally, approaches such as {\tt metalibm}~\cite{7203797} for math function code generation could be leveraged to produce specialized libraries.

To conclude, the work presented in this paper shows promising results on co-designing mathematical libraries from application profiles. 
One weakness of the approach is that the profile is data-input dependent; further experiments on a larger set of use-cases will be done to demonstrate the generalization of the approach and how one can deal with the speculative aspect of profile-guided optimization for math libraries.

\bibliographystyle{IEEEtran}
\bibliography{biblio,mistoan}

\end{document}

%% file: vprec_libm_overview.tex
\begin{center}
	\begin{figure}[htb]
		\scalebox{0.67}{
			\begin{tikzpicture}[node distance=2em, auto]
			
			\node [process, minimum width=6em, minimum height=5em, align=center] (binary) {Binary\\\\\small\texttt{0101000}\\\small\texttt{0111011}\\\texttt{$\ldots$}};
			\node [doc, below = 2em of binary, minimum width=6em, minimum height=3em, align=center] (profile) {Profile};
			\node [doc, below = 1em of profile, minimum width=4.5em, minimum height=3em, align=center] (optConfig) {Optimized\\Precisions};
			
			\node [process, left = 6em of binary, minimum width=4em, minimum height=3em, align=center] (execute) {Execute};
			\node [below = 2em of execute, minimum width=4em, align=center] (ldpreload) {\texttt{LD\_PRELOAD}};
			\node [dataStructure, below = 2em of ldpreload, text width=4em, minimum width=4em, minimum height=2em, align=center] (vplibm) {\texttt{VPREC\\libm}};
			
			\node [process, right = 6em of profile, minimum width=4em, minimum height=3em, align=center] (explore) {Dichotomic\\Exploration};
			\node [doc, below = 2em of explore, minimum width=4em, minimum height=3em, align=center] (config) {Config.};
			
			\node [process, right = 13em of binary, minimum width=4em, minimum height=3em, align=center] (execute2) {Execute};
			\node [below = 2em of execute2, minimum width=4em, align=center] (ldpreload2) {\texttt{LD\_PRELOAD}};
			\node [dataStructure, below = 2em of ldpreload2, text width=4em, minimum width=4em, minimum height=2em, align=center] (vplibm2) {\texttt{VPREC\\libm}};
			
			\node [rectangle, draw, rounded corners, dashed, above left = -14em and -6em of execute, minimum width=8em, minimum height=16em, align=center] (boxProfile) {};
			\node [below right = -1.5em and -5.75em of boxProfile] (boxProfileText) {\textit{First Profile}};
			\node [rectangle, draw, rounded corners, dashed, above left = -16em and -6em of execute2, minimum width=15em, minimum height=18em, align=center] (boxExplore) {};
			\node [below right = -1.5em and -5.5em of boxExplore] (boxExploreText) {\textit{Exploration}};
			
			\path [line] (binary.west) -- (execute.east);
			\path [line] (binary.east) -- (execute2.west);
			
			\path [line] (execute) -- (profile.west);
			
			\path [dottedLine] (ldpreload.north) -- (execute.south);
			\path [dottedLineSimple] (vplibm.north) -- (ldpreload.south);
			
			\path [line] (explore) -- (optConfig.east);
			\path [line] (explore.south) -- (config.north);
			\path [line] (config.east) -| (vplibm2.south);
			
			\path [line] (explore) edge [loop above,looseness=2, min distance=3em, out=55, in=115] (explore);
			
			\path [line] (execute2) -- (explore);
			
			\path [dottedLine] (ldpreload2.north) -- (execute2.south);
			\path [dottedLineSimple] (vplibm2.north) -- (ldpreload2.south);		
			
			\end{tikzpicture}
		}
		\caption{\texttt{VPREC-libm} optimization process overview}
		\label{fig:vprec-libm_overview}
	\end{figure}
\end{center}

%% file: all_sat_freq.tex
\begin{tikzpicture}
	\begin{axis}[
		ybar = -7pt,
		y tick label style = {/pgf/number format/set thousands separator=,},
		width = 1.0\textwidth,
		height = 0.15\textheight,
		symbolic x coords = { 0\_FABS, 1\_COS, 2\_SIN, 3\_FABS, 4\_POW, 5\_POW, 6\_FMOD, 7\_FMOD, 8\_FMOD, 9\_FMOD, 10\_FMOD, 11\_ATAN2, 12\_ATAN2, 13\_FMOD, 14\_FMOD, 15\_FMOD, 16\_SQRT, 17\_SIN, 18\_COS, 19\_SIN, 20\_COS, 21\_SIN, 22\_SQRT, 23\_COS, 24\_SIN, 25\_COS, 26\_SIN, 27\_COS, 28\_SQRT, 29\_SQRT, 30\_ACOS, 31\_SIN, 32\_FLOOR, 33\_SQRT, 34\_SQRT, 35\_FLOOR, 36\_SQRT, 37\_SQRT, 38\_FABS, 39\_SQRT, 40\_SQRT, 41\_SQRT, 42\_FABS, 43\_ACOS, 44\_FABS, 45\_SQRT, 46\_SQRT, 47\_FLOOR, 48\_ACOS, 49\_COS },
		bar direction = y,
		bar width = 7pt,
		tick align = center,
		enlargelimits = 0.01,
		xtick = data,
		xticklabels={,,},
		ymin=0,ymax=3601,
		ylabel = Number of Calls,
		y label style = {scale=1.5, transform shape},
		x tick label style = {rotate=45,anchor=east},
		nodes near coords,
		nodes near coords align = {vertical},
		nodes near coords style = {yshift=1.5,rotate=90,anchor=east,/pgf/number format/set thousands separator={}},
		legend style = {at={(1.14,0.75)}, anchor=east, legend columns=-1, align=center},
	]
		\addplot[draw=teal, fill=cyan!30] coordinates { (0\_FABS, 3187) (1\_COS, 2549) (2\_SIN, 2549) (3\_FABS, 2549) (4\_POW, 1280) (5\_POW, 1280) (6\_FMOD, 1276) (7\_FMOD, 1276) (8\_FMOD, 1276) (9\_FMOD, 1276) (10\_FMOD, 1276) (11\_ATAN2, 1274) (12\_ATAN2, 1268) (13\_FMOD, 1268) (14\_FMOD, 1268) (15\_FMOD, 974) (16\_SQRT, 640) (17\_SIN, 638) (18\_COS, 638) (19\_SIN, 638) (20\_COS, 638) (21\_SIN, 637) (22\_SQRT, 637) (23\_COS, 637) (24\_SIN, 637) (25\_COS, 637) (26\_SIN, 637) (27\_COS, 637) (28\_SQRT, 637) (29\_SQRT, 637) (30\_ACOS, 634) (31\_SIN, 634) (32\_FLOOR, 634) (33\_SQRT, 634) (34\_SQRT, 634) (35\_FLOOR, 634) (36\_SQRT, 634) (37\_SQRT, 634) (38\_FABS, 634) (39\_SQRT, 634) (40\_SQRT, 634) (41\_SQRT, 634) (42\_FABS, 634) (43\_ACOS, 634) (44\_FABS, 634) (45\_SQRT, 634) (46\_SQRT, 634) (47\_FLOOR, 634) (48\_ACOS, 634) (49\_COS, 634)};
		\legend{\small Original\\\small and\\\small Optimized}
	\end{axis}
\end{tikzpicture}

%% file: all_sat_ranges2.tex
\begin{tikzpicture}
	\begin{axis}[
		y tick label style = {/pgf/number format/set thousands separator=,},
		enlargelimits = 0.01,
		width = 1.0\textwidth,
		height = 0.155\textheight,
		boxplot/draw direction = y,
		xtick = { 1, 2, 3, 4, 5, 6, 7, 8, 9, 10, 11, 12, 13, 14, 15, 16, 17, 18, 19, 20, 21, 22, 23, 24, 25, 26, 27, 28, 29, 30, 31, 32, 33, 34, 35, 36, 37, 38, 39, 40, 41, 42, 43, 44, 45, 46, 47, 48, 49, 50 },
		xticklabels = { 0\_FABS, 1\_COS, 2\_SIN, 3\_FABS, 4\_POW, 5\_POW, 6\_FMOD, 7\_FMOD, 8\_FMOD, 9\_FMOD, 10\_FMOD, 11\_ATAN2, 12\_ATAN2, 13\_FMOD, 14\_FMOD, 15\_FMOD, 16\_SQRT, 17\_SIN, 18\_COS, 19\_SIN, 20\_COS, 21\_SIN, 22\_SQRT, 23\_COS, 24\_SIN, 25\_COS, 26\_SIN, 27\_COS, 28\_SQRT, 29\_SQRT, 30\_ACOS, 31\_SIN, 32\_FLOOR, 33\_SQRT, 34\_SQRT, 35\_FLOOR, 36\_SQRT, 37\_SQRT, 38\_FABS, 39\_SQRT, 40\_SQRT, 41\_SQRT, 42\_FABS, 43\_ACOS, 44\_FABS, 45\_SQRT, 46\_SQRT, 47\_FLOOR, 48\_ACOS, 49\_COS },
		ymajorgrids,
		xmajorgrids,
		ymode = log,
		log basis y = 2,
		ymin=2^-36,ymax=2^36,
		ylabel = Input interval\\ size (log),
		y label style = {scale=1.5, transform shape, align=center},
		tick align = center,
		xticklabels={,,},
		x tick label style = {rotate=45,anchor=east},
		legend style = {at={(1.14,0.75)}, anchor=east, legend columns=1},
		cycle list={{olive}},
	]
		\legend{\small Optimized}
		\addplot+ [
		draw=olive, fill=lime!40,
		boxplot prepared = { box extend=0.75, draw position=1, lower whisker=1.0, lower quartile=1.0, median=, upper quartile=9999.9, upper whisker=9999.9, average= },
		] coordinates {};
		\addplot+ [
		draw=olive, fill=lime!40,
		boxplot prepared = { box extend=0.75, draw position=2, lower whisker=0.002880508, lower quartile=0.002880508, median=, upper quartile=6.691296, upper whisker=6.691296, average= },
		] coordinates {};
		\addplot+ [
		draw=olive, fill=lime!40,
		boxplot prepared = { box extend=0.75, draw position=3, lower whisker=0.002880508, lower quartile=0.002880508, median=, upper quartile=6.691296, upper whisker=6.691296, average= },
		] coordinates {};
		\addplot+ [
		draw=olive, fill=lime!40,
		boxplot prepared = { box extend=0.75, draw position=4, lower whisker=1.0, lower quartile=1.0, median=, upper quartile=7.271535, upper whisker=7.271535, average= },
		] coordinates {};
		\addplot+ [
		draw=olive, fill=lime!40,
		boxplot prepared = { box extend=0.75, draw position=5, lower whisker=0.6666667, lower quartile=0.6666667, median=, upper quartile=233.1853, upper whisker=233.1853, average= },
		] coordinates {};
		\addplot+ [
		draw=olive, fill=lime!40,
		boxplot prepared = { box extend=0.75, draw position=6, lower whisker=0.9956389, lower quartile=0.9956389, median=, upper quartile=37.88501, upper whisker=37.88501, average= },
		] coordinates {};
		\addplot+ [
		draw=olive, fill=lime!40,
		boxplot prepared = { box extend=0.75, draw position=7, lower whisker=0.01076291, lower quartile=0.01076291, median=, upper quartile=13.04557, upper whisker=13.04557, average= },
		] coordinates {};
		\addplot+ [
		draw=olive, fill=lime!40,
		boxplot prepared = { box extend=0.75, draw position=8, lower whisker=0.03852097, lower quartile=0.03852097, median=, upper quartile=1998.92, upper whisker=1998.92, average= },
		] coordinates {};
		\addplot+ [
		draw=olive, fill=lime!40,
		boxplot prepared = { box extend=0.75, draw position=9, lower whisker=6.007871e-06, lower quartile=6.007871e-06, median=, upper quartile=6.283185, upper whisker=6.283185, average= },
		] coordinates {};
		\addplot+ [
		draw=olive, fill=lime!40,
		boxplot prepared = { box extend=0.75, draw position=10, lower whisker=0.2407195, lower quartile=0.2407195, median=, upper quartile=6.837239, upper whisker=6.837239, average= },
		] coordinates {};
		\addplot+ [
		draw=olive, fill=lime!40,
		boxplot prepared = { box extend=0.75, draw position=11, lower whisker=0.008635214, lower quartile=0.008635214, median=, upper quartile=6.283185, upper whisker=6.283185, average= },
		] coordinates {};
		\addplot+ [
		draw=olive, fill=lime!40,
		boxplot prepared = { box extend=0.75, draw position=12, lower whisker=8.881252e-05, lower quartile=8.881252e-05, median=, upper quartile=1.0, upper whisker=1.0, average= },
		] coordinates {};
		\addplot+ [
		draw=olive, fill=lime!40,
		boxplot prepared = { box extend=0.75, draw position=13, lower whisker=0.0002601703, lower quartile=0.0002601703, median=, upper quartile=1.0, upper whisker=1.0, average= },
		] coordinates {};
		\addplot+ [
		draw=olive, fill=lime!40,
		boxplot prepared = { box extend=0.75, draw position=14, lower whisker=0.0002601703, lower quartile=0.0002601703, median=, upper quartile=6.283185, upper whisker=6.283185, average= },
		] coordinates {};
		\addplot+ [
		draw=olive, fill=lime!40,
		boxplot prepared = { box extend=0.75, draw position=15, lower whisker=0.0002601609, lower quartile=0.0002601609, median=, upper quartile=6.283185, upper whisker=6.283185, average= },
		] coordinates {};
		\addplot+ [
		draw=olive, fill=lime!40,
		boxplot prepared = { box extend=0.75, draw position=16, lower whisker=0.1213336, lower quartile=0.1213336, median=, upper quartile=8069.913, upper whisker=8069.913, average= },
		] coordinates {};
		\addplot+ [
		draw=olive, fill=lime!40,
		boxplot prepared = { box extend=0.75, draw position=17, lower whisker=650943.0, lower quartile=650943.0, median=, upper quartile=650943.0, upper whisker=650943.0, average= },
		] coordinates {};
		\addplot+ [
		draw=olive, fill=lime!40,
		boxplot prepared = { box extend=0.75, draw position=18, lower whisker=1.528713e-07, lower quartile=1.528713e-07, median=, upper quartile=1.717898, upper whisker=1.717898, average= },
		] coordinates {};
		\addplot+ [
		draw=olive, fill=lime!40,
		boxplot prepared = { box extend=0.75, draw position=19, lower whisker=1.528713e-07, lower quartile=1.528713e-07, median=, upper quartile=1.717898, upper whisker=1.717898, average= },
		] coordinates {};
		\addplot+ [
		draw=olive, fill=lime!40,
		boxplot prepared = { box extend=0.75, draw position=20, lower whisker=0.03776485, lower quartile=0.03776485, median=, upper quartile=6.024962, upper whisker=6.024962, average= },
		] coordinates {};
		\addplot+ [
		draw=olive, fill=lime!40,
		boxplot prepared = { box extend=0.75, draw position=21, lower whisker=0.03776485, lower quartile=0.03776485, median=, upper quartile=6.024962, upper whisker=6.024962, average= },
		] coordinates {};
		\addplot+ [
		draw=olive, fill=lime!40,
		boxplot prepared = { box extend=0.75, draw position=22, lower whisker=0.002493096, lower quartile=0.002493096, median=, upper quartile=3.133251, upper whisker=3.133251, average= },
		] coordinates {};
		\addplot+ [
		draw=olive, fill=lime!40,
		boxplot prepared = { box extend=0.75, draw position=23, lower whisker=0.009994303, lower quartile=0.009994303, median=, upper quartile=6.673346, upper whisker=6.673346, average= },
		] coordinates {};
		\addplot+ [
		draw=olive, fill=lime!40,
		boxplot prepared = { box extend=0.75, draw position=24, lower whisker=5.85877e-05, lower quartile=5.85877e-05, median=, upper quartile=1.717991, upper whisker=1.717991, average= },
		] coordinates {};
		\addplot+ [
		draw=olive, fill=lime!40,
		boxplot prepared = { box extend=0.75, draw position=25, lower whisker=5.85877e-05, lower quartile=5.85877e-05, median=, upper quartile=1.717991, upper whisker=1.717991, average= },
		] coordinates {};
		\addplot+ [
		draw=olive, fill=lime!40,
		boxplot prepared = { box extend=0.75, draw position=26, lower whisker=5.091377e-05, lower quartile=5.091377e-05, median=, upper quartile=6.317846, upper whisker=6.317846, average= },
		] coordinates {};
		\addplot+ [
		draw=olive, fill=lime!40,
		boxplot prepared = { box extend=0.75, draw position=27, lower whisker=5.091377e-05, lower quartile=5.091377e-05, median=, upper quartile=6.317846, upper whisker=6.317846, average= },
		] coordinates {};
		\addplot+ [
		draw=olive, fill=lime!40,
		boxplot prepared = { box extend=0.75, draw position=28, lower whisker=0.002493096, lower quartile=0.002493096, median=, upper quartile=3.133251, upper whisker=3.133251, average= },
		] coordinates {};
		\addplot+ [
		draw=olive, fill=lime!40,
		boxplot prepared = { box extend=0.75, draw position=29, lower whisker=0.9956389, lower quartile=0.9956389, median=, upper quartile=37.88501, upper whisker=37.88501, average= },
		] coordinates {};
		\addplot+ [
		draw=olive, fill=lime!40,
		boxplot prepared = { box extend=0.75, draw position=30, lower whisker=0.004102251, lower quartile=0.004102251, median=, upper quartile=1.0, upper whisker=1.0, average= },
		] coordinates {};
		\addplot+ [
		draw=olive, fill=lime!40,
		boxplot prepared = { box extend=0.75, draw position=31, lower whisker=0.1065937, lower quartile=0.1065937, median=, upper quartile=1.0, upper whisker=1.0, average= },
		] coordinates {};
		\addplot+ [
		draw=olive, fill=lime!40,
		boxplot prepared = { box extend=0.75, draw position=32, lower whisker=0.0002601703, lower quartile=0.0002601703, median=, upper quartile=3.138574, upper whisker=3.138574, average= },
		] coordinates {};
		\addplot+ [
		draw=olive, fill=lime!40,
		boxplot prepared = { box extend=0.75, draw position=33, lower whisker=1.44064e-08, lower quartile=1.44064e-08, median=, upper quartile=59.99999, upper whisker=59.99999, average= },
		] coordinates {};
		\addplot+ [
		draw=olive, fill=lime!40,
		boxplot prepared = { box extend=0.75, draw position=34, lower whisker=0.4626929, lower quartile=0.4626929, median=, upper quartile=106.0569, upper whisker=106.0569, average= },
		] coordinates {};
		\addplot+ [
		draw=olive, fill=lime!40,
		boxplot prepared = { box extend=0.75, draw position=35, lower whisker=27307170.0, lower quartile=27307170.0, median=, upper quartile=16965870000.0, upper whisker=16965870000.0, average= },
		] coordinates {};
		\addplot+ [
		draw=olive, fill=lime!40,
		boxplot prepared = { box extend=0.75, draw position=36, lower whisker=80.62846, lower quartile=80.62846, median=, upper quartile=109.5019, upper whisker=109.5019, average= },
		] coordinates {};
		\addplot+ [
		draw=olive, fill=lime!40,
		boxplot prepared = { box extend=0.75, draw position=37, lower whisker=1.814772e-11, lower quartile=1.814772e-11, median=, upper quartile=0.9971281, upper whisker=0.9971281, average= },
		] coordinates {};
		\addplot+ [
		draw=olive, fill=lime!40,
		boxplot prepared = { box extend=0.75, draw position=38, lower whisker=57.69213, lower quartile=57.69213, median=, upper quartile=7057322000.0, upper whisker=7057322000.0, average= },
		] coordinates {};
		\addplot+ [
		draw=olive, fill=lime!40,
		boxplot prepared = { box extend=0.75, draw position=39, lower whisker=0.0008959689, lower quartile=0.0008959689, median=, upper quartile=1.0, upper whisker=1.0, average= },
		] coordinates {};
		\addplot+ [
		draw=olive, fill=lime!40,
		boxplot prepared = { box extend=0.75, draw position=40, lower whisker=1.814772e-11, lower quartile=1.814772e-11, median=, upper quartile=0.9971281, upper whisker=0.9971281, average= },
		] coordinates {};
		\addplot+ [
		draw=olive, fill=lime!40,
		boxplot prepared = { box extend=0.75, draw position=41, lower whisker=40738250.0, lower quartile=40738250.0, median=, upper quartile=48618900000.0, upper whisker=48618900000.0, average= },
		] coordinates {};
		\addplot+ [
		draw=olive, fill=lime!40,
		boxplot prepared = { box extend=0.75, draw position=42, lower whisker=40738250.0, lower quartile=40738250.0, median=, upper quartile=48618900000.0, upper whisker=48618900000.0, average= },
		] coordinates {};
		\addplot+ [
		draw=olive, fill=lime!40,
		boxplot prepared = { box extend=0.75, draw position=43, lower whisker=0.8253285, lower quartile=0.8253285, median=, upper quartile=31.25981, upper whisker=31.25981, average= },
		] coordinates {};
		\addplot+ [
		draw=olive, fill=lime!40,
		boxplot prepared = { box extend=0.75, draw position=44, lower whisker=0.02352192, lower quartile=0.02352192, median=, upper quartile=1.0, upper whisker=1.0, average= },
		] coordinates {};
		\addplot+ [
		draw=olive, fill=lime!40,
		boxplot prepared = { box extend=0.75, draw position=45, lower whisker=1.423602, lower quartile=1.423602, median=, upper quartile=3.141534, upper whisker=3.141534, average= },
		] coordinates {};
		\addplot+ [
		draw=olive, fill=lime!40,
		boxplot prepared = { box extend=0.75, draw position=46, lower whisker=57.69213, lower quartile=57.69213, median=, upper quartile=7057322000.0, upper whisker=7057322000.0, average= },
		] coordinates {};
		\addplot+ [
		draw=olive, fill=lime!40,
		boxplot prepared = { box extend=0.75, draw position=47, lower whisker=1.814772e-11, lower quartile=1.814772e-11, median=, upper quartile=0.9971281, upper whisker=0.9971281, average= },
		] coordinates {};
		\addplot+ [
		draw=olive, fill=lime!40,
		boxplot prepared = { box extend=0.75, draw position=48, lower whisker=0.03517779, lower quartile=0.03517779, median=, upper quartile=24.0, upper whisker=24.0, average= },
		] coordinates {};
		\addplot+ [
		draw=olive, fill=lime!40,
		boxplot prepared = { box extend=0.75, draw position=49, lower whisker=0.0008959689, lower quartile=0.0008959689, median=, upper quartile=1.0, upper whisker=1.0, average= },
		] coordinates {};
		\addplot+ [
		draw=olive, fill=lime!40,
		boxplot prepared = { box extend=0.75, draw position=50, lower whisker=0.0006174048, lower quartile=0.0006174048, median=, upper quartile=6.282925, upper whisker=6.282925, average= },
		] coordinates {};
	\end{axis}
\end{tikzpicture}

%% file: all_sat_prec.tex
\begin{tikzpicture}
	\begin{axis}[
		ybar = -7pt,
		width = 1.0\textwidth,
		height = 0.175\textheight,
		symbolic x coords = { 0\_FABS, 1\_COS, 2\_SIN, 3\_FABS, 4\_POW, 5\_POW, 6\_FMOD, 7\_FMOD, 8\_FMOD, 9\_FMOD, 10\_FMOD, 11\_ATAN2, 12\_ATAN2, 13\_FMOD, 14\_FMOD, 15\_FMOD, 16\_SQRT, 17\_SIN, 18\_COS, 19\_SIN, 20\_COS, 21\_SIN, 22\_SQRT, 23\_COS, 24\_SIN, 25\_COS, 26\_SIN, 27\_COS, 28\_SQRT, 29\_SQRT, 30\_ACOS, 31\_SIN, 32\_FLOOR, 33\_SQRT, 34\_SQRT, 35\_FLOOR, 36\_SQRT, 37\_SQRT, 38\_FABS, 39\_SQRT, 40\_SQRT, 41\_SQRT, 42\_FABS, 43\_ACOS, 44\_FABS, 45\_SQRT, 46\_SQRT, 47\_FLOOR, 48\_ACOS, 49\_COS },
		bar direction = y,
		bar width = 7pt,
		tick align = center,
		enlargelimits = 0.01,
		xtick = data,
		ymin=0,ymax=60,
		ylabel = Vprec-libm output\\ precision (bits),
		y label style = {scale=1.5, transform shape, align=center},
		extra y ticks={23, 28},
		x tick label style = {rotate=45,anchor=east},
		nodes near coords,
		nodes near coords align = {vertical},
		nodes near coords style = {/pgf/number format/set thousands separator={}},
		legend style = {at={(1.14,0.75)}, anchor=east, legend columns=1},
	]
		\addplot[draw=orange, fill=pink!40] coordinates { (0\_FABS, 52) (1\_COS, 52) (2\_SIN, 52) (3\_FABS, 52) (4\_POW, 52) (5\_POW, 52) (6\_FMOD, 52) (7\_FMOD, 52) (8\_FMOD, 52) (9\_FMOD, 52) (10\_FMOD, 52) (11\_ATAN2, 52) (12\_ATAN2, 52) (13\_FMOD, 52) (14\_FMOD, 52) (15\_FMOD, 52) (16\_SQRT, 52) (17\_SIN, 52) (18\_COS, 52) (19\_SIN, 52) (20\_COS, 52) (21\_SIN, 52) (22\_SQRT, 52) (23\_COS, 52) (24\_SIN, 52) (25\_COS, 52) (26\_SIN, 52) (27\_COS, 52) (28\_SQRT, 52) (29\_SQRT, 52) (30\_ACOS, 52) (31\_SIN, 52) (32\_FLOOR, 52) (33\_SQRT, 52) (34\_SQRT, 52) (35\_FLOOR, 52) (36\_SQRT, 52) (37\_SQRT, 52) (38\_FABS, 52) (39\_SQRT, 52) (40\_SQRT, 52) (41\_SQRT, 52) (42\_FABS, 52) (43\_ACOS, 52) (44\_FABS, 52) (45\_SQRT, 52) (46\_SQRT, 52) (47\_FLOOR, 52) (48\_ACOS, 52) (49\_COS, 52)};
		\addplot[draw=olive, fill=lime!40, nodes near coords style = {yshift=0}] coordinates { (0\_FABS, 2) (1\_COS, 50) (2\_SIN, 50) (3\_FABS, 0) (4\_POW, 49) (5\_POW, 29) (6\_FMOD, 52) (7\_FMOD, 49) (8\_FMOD, 52) (9\_FMOD, 52) (10\_FMOD, 51) (11\_ATAN2, 50) (12\_ATAN2, 31) (13\_FMOD, 0) (14\_FMOD, 33) (15\_FMOD, 49) (16\_SQRT, 45) (17\_SIN, 30) (18\_COS, 33) (19\_SIN, 48) (20\_COS, 51) (21\_SIN, 49) (22\_SQRT, 45) (23\_COS, 46) (24\_SIN, 49) (25\_COS, 48) (26\_SIN, 48) (27\_COS, 50) (28\_SQRT, 47) (29\_SQRT, 44) (30\_ACOS, 32) (31\_SIN, 28) (32\_FLOOR, 5) (33\_SQRT, 45) (34\_SQRT, 38) (35\_FLOOR, 6) (36\_SQRT, 41) (37\_SQRT, 41) (38\_FABS, 0) (39\_SQRT, 29) (40\_SQRT, 45) (41\_SQRT, 37) (42\_FABS, 0) (43\_ACOS, 29) (44\_FABS, 0) (45\_SQRT, 39) (46\_SQRT, 37) (47\_FLOOR, 4) (48\_ACOS, 33) (49\_COS, 34)};
		\addplot[orange,sharp plot,nodes near coords={},update limits=false,shorten >=-3mm,shorten <=-3mm] coordinates {(0\_FABS,23) (49\_COS,23)};
		\addplot[teal,sharp plot,nodes near coords={},update limits=false,shorten >=-3mm,shorten <=-3mm] coordinates {(0\_FABS,28) (49\_COS,28)};
		\legend{\small Original, \small Optimized}
	\end{axis}
\end{tikzpicture}

%% file: 5_sat_freq.tex
\begin{tikzpicture}
	\begin{axis}[
		ybar = -7pt,
		y tick label style = {/pgf/number format/set thousands separator=,},
		width = 1.0\textwidth,
		height = 0.15\textheight,
		symbolic x coords = { 0\_FABS, 1\_COS, 2\_FABS, 3\_SIN, 4\_FMOD, 5\_ATAN2, 6\_ATAN2, 7\_FMOD, 8\_FMOD, 9\_FMOD, 10\_POW, 11\_POW, 12\_FMOD, 13\_FMOD, 14\_FMOD, 15\_ACOS, 16\_SQRT, 17\_FLOOR, 18\_COS, 19\_ACOS, 20\_FABS, 21\_SQRT, 22\_FLOOR, 23\_COS, 24\_COS, 25\_FABS, 26\_SIN, 27\_SQRT, 28\_SQRT, 29\_FLOOR, 30\_COS, 31\_COS, 32\_FABS, 33\_ACOS, 34\_SQRT, 35\_SIN, 36\_SQRT, 37\_SQRT, 38\_SIN, 39\_ACOS, 40\_SQRT, 41\_FLOOR, 42\_COS, 43\_SIN, 44\_FABS, 45\_SIN, 46\_SQRT, 47\_SIN, 48\_SQRT, 49\_FLOOR },
		bar direction = y,
		bar width = 7pt,
		tick align = center,
		enlargelimits = 0.01,
		xtick = data,
		xticklabels={,,},
		ymin=0,ymax=65,
		ylabel = Number of Calls,
		y label style = {scale=1.5, transform shape},
		x tick label style = {rotate=45,anchor=east},
		nodes near coords,
		nodes near coords align = {vertical},
		nodes near coords style = {yshift=1.5,rotate=90,anchor=east,/pgf/number format/set thousands separator={}},
		legend style = {at={(1.14,0.75)}, anchor=east, legend columns=-1, align=center},
	]
		\addplot[draw=teal, fill=cyan!30] coordinates { (0\_FABS, 58) (1\_COS, 46) (2\_FABS, 46) (3\_SIN, 46) (4\_FMOD, 24) (5\_ATAN2, 24) (6\_ATAN2, 24) (7\_FMOD, 24) (8\_FMOD, 24) (9\_FMOD, 24) (10\_POW, 24) (11\_POW, 24) (12\_FMOD, 24) (13\_FMOD, 24) (14\_FMOD, 24) (15\_ACOS, 12) (16\_SQRT, 12) (17\_FLOOR, 12) (18\_COS, 12) (19\_ACOS, 12) (20\_FABS, 12) (21\_SQRT, 12) (22\_FLOOR, 12) (23\_COS, 12) (24\_COS, 12) (25\_FABS, 12) (26\_SIN, 12) (27\_SQRT, 12) (28\_SQRT, 12) (29\_FLOOR, 12) (30\_COS, 12) (31\_COS, 12) (32\_FABS, 12) (33\_ACOS, 12) (34\_SQRT, 12) (35\_SIN, 12) (36\_SQRT, 12) (37\_SQRT, 12) (38\_SIN, 12) (39\_ACOS, 12) (40\_SQRT, 12) (41\_FLOOR, 12) (42\_COS, 12) (43\_SIN, 12) (44\_FABS, 12) (45\_SIN, 12) (46\_SQRT, 12) (47\_SIN, 12) (48\_SQRT, 12) (49\_FLOOR, 12)};
		\legend{\small Original\\\small and\\\small Optimized}
	\end{axis}
\end{tikzpicture}

%% file: 5_sat_ranges2.tex
\begin{tikzpicture}
	\begin{axis}[
		y tick label style = {/pgf/number format/set thousands separator=,},
		enlargelimits = 0.01,
		width = 1.0\textwidth,
		height = 0.15\textheight,
		boxplot/draw direction = y,
		xtick = { 1, 2, 3, 4, 5, 6, 7, 8, 9, 10, 11, 12, 13, 14, 15, 16, 17, 18, 19, 20, 21, 22, 23, 24, 25, 26, 27, 28, 29, 30, 31, 32, 33, 34, 35, 36, 37, 38, 39, 40, 41, 42, 43, 44, 45, 46, 47, 48, 49, 50 },
		xticklabels = { 0\_FABS, 1\_COS, 2\_SIN, 3\_FABS, 4\_FMOD, 5\_ATAN2, 6\_ATAN2, 7\_FMOD, 8\_FMOD, 9\_FMOD, 10\_POW, 11\_POW, 12\_FMOD, 13\_FMOD, 14\_FMOD, 15\_ACOS, 16\_SQRT, 17\_FLOOR, 18\_COS, 19\_ACOS, 20\_FABS, 21\_SQRT, 22\_FLOOR, 23\_COS, 24\_COS, 25\_FABS, 26\_SIN, 27\_SQRT, 28\_SQRT, 29\_FLOOR, 30\_COS, 31\_COS, 32\_FABS, 33\_ACOS, 34\_SQRT, 35\_SIN, 36\_SQRT, 37\_SQRT, 38\_SIN, 39\_ACOS, 40\_SQRT, 41\_FLOOR, 42\_COS, 43\_SIN, 44\_FABS, 45\_SIN, 46\_SQRT, 47\_SIN, 48\_SQRT, 49\_FLOOR },
		ymajorgrids,
		xmajorgrids,
		ymode = log,
		log basis y = 2,
		ymin=2^-36,ymax=2^36,
		ylabel = Input interval\\ size (log),
		y label style = {scale=1.5, transform shape, align=center},
		tick align = center,
		xticklabels={,,},
		x tick label style = {rotate=45,anchor=east},
		legend style = {at={(1.14,0.75)}, anchor=east, legend columns=1},
		cycle list={{olive}},
	]
		\legend{\small Optimized}
		\addplot+ [
		draw=olive, fill=lime!40,
		boxplot prepared = { box extend=0.75, draw position=1, lower whisker=1.0, lower quartile=1.0, median=, upper quartile=9999.9, upper whisker=9999.9, average= },
		] coordinates {};
		\addplot+ [
		draw=olive, fill=lime!40,
		boxplot prepared = { box extend=0.75, draw position=2, lower whisker=0.420172, lower quartile=0.420172, median=, upper quartile=5.902397, upper whisker=5.902397, average= },
		] coordinates {};
		\addplot+ [
		draw=olive, fill=lime!40,
		boxplot prepared = { box extend=0.75, draw position=3, lower whisker=0.420172, lower quartile=0.420172, median=, upper quartile=5.902397, upper whisker=5.902397, average= },
		] coordinates {};
		\addplot+ [
		draw=olive, fill=lime!40,
		boxplot prepared = { box extend=0.75, draw position=4, lower whisker=1.0, lower quartile=1.0, median=, upper quartile=0.1874644, upper whisker=0.1874644, average= },
		] coordinates {};
		\addplot+ [
		draw=olive, fill=lime!40,
		boxplot prepared = { box extend=0.75, draw position=5, lower whisker=5.925805, lower quartile=5.925805, median=, upper quartile=6.283185, upper whisker=6.283185, average= },
		] coordinates {};
		\addplot+ [
		draw=olive, fill=lime!40,
		boxplot prepared = { box extend=0.75, draw position=6, lower whisker=0.03321011, lower quartile=0.03321011, median=, upper quartile=0.9994484, upper whisker=0.9994484, average= },
		] coordinates {};
		\addplot+ [
		draw=olive, fill=lime!40,
		boxplot prepared = { box extend=0.75, draw position=7, lower whisker=0.02187485, lower quartile=0.02187485, median=, upper quartile=0.9997607, upper whisker=0.9997607, average= },
		] coordinates {};
		\addplot+ [
		draw=olive, fill=lime!40,
		boxplot prepared = { box extend=0.75, draw position=8, lower whisker=6.283185, lower quartile=6.283185, median=, upper quartile=11.84906, upper whisker=11.84906, average= },
		] coordinates {};
		\addplot+ [
		draw=olive, fill=lime!40,
		boxplot prepared = { box extend=0.75, draw position=9, lower whisker=0.02702795, lower quartile=0.02702795, median=, upper quartile=6.283185, upper whisker=6.283185, average= },
		] coordinates {};
		\addplot+ [
		draw=olive, fill=lime!40,
		boxplot prepared = { box extend=0.75, draw position=10, lower whisker=6.283185, lower quartile=6.283185, median=, upper quartile=216.3193, upper whisker=216.3193, average= },
		] coordinates {};
		\addplot+ [
		draw=olive, fill=lime!40,
		boxplot prepared = { box extend=0.75, draw position=11, lower whisker=1.3539, lower quartile=1.3539, median=, upper quartile=1.5, upper whisker=1.5, average= },
		] coordinates {};
		\addplot+ [
		draw=olive, fill=lime!40,
		boxplot prepared = { box extend=0.75, draw position=12, lower whisker=0.6666667, lower quartile=0.6666667, median=, upper quartile=1.57536, upper whisker=1.57536, average= },
		] coordinates {};
		\addplot+ [
		draw=olive, fill=lime!40,
		boxplot prepared = { box extend=0.75, draw position=13, lower whisker=5.809962, lower quartile=5.809962, median=, upper quartile=6.283185, upper whisker=6.283185, average= },
		] coordinates {};
		\addplot+ [
		draw=olive, fill=lime!40,
		boxplot prepared = { box extend=0.75, draw position=14, lower whisker=0.03321622, lower quartile=0.03321622, median=, upper quartile=6.283185, upper whisker=6.283185, average= },
		] coordinates {};
		\addplot+ [
		draw=olive, fill=lime!40,
		boxplot prepared = { box extend=0.75, draw position=15, lower whisker=0.420172, lower quartile=0.420172, median=, upper quartile=6.283185, upper whisker=6.283185, average= },
		] coordinates {};
		\addplot+ [
		draw=olive, fill=lime!40,
		boxplot prepared = { box extend=0.75, draw position=16, lower whisker=0.8262865, lower quartile=0.8262865, median=, upper quartile=0.8265337, upper whisker=0.8265337, average= },
		] coordinates {};
		\addplot+ [
		draw=olive, fill=lime!40,
		boxplot prepared = { box extend=0.75, draw position=17, lower whisker=3321980000.0, lower quartile=3321980000.0, median=, upper quartile=3324854000.0, upper whisker=3324854000.0, average= },
		] coordinates {};
		\addplot+ [
		draw=olive, fill=lime!40,
		boxplot prepared = { box extend=0.75, draw position=18, lower whisker=100.4902, lower quartile=100.4902, median=, upper quartile=100.4977, upper whisker=100.4977, average= },
		] coordinates {};
		\addplot+ [
		draw=olive, fill=lime!40,
		boxplot prepared = { box extend=0.75, draw position=19, lower whisker=0.3937781, lower quartile=0.3937781, median=, upper quartile=3.119729, upper whisker=3.119729, average= },
		] coordinates {};
		\addplot+ [
		draw=olive, fill=lime!40,
		boxplot prepared = { box extend=0.75, draw position=20, lower whisker=0.8898743, lower quartile=0.8898743, median=, upper quartile=0.9674455, upper whisker=0.9674455, average= },
		] coordinates {};
		\addplot+ [
		draw=olive, fill=lime!40,
		boxplot prepared = { box extend=0.75, draw position=21, lower whisker=23.07438, lower quartile=23.07438, median=, upper quartile=23.08829, upper whisker=23.08829, average= },
		] coordinates {};
		\addplot+ [
		draw=olive, fill=lime!40,
		boxplot prepared = { box extend=0.75, draw position=22, lower whisker=0.03424871, lower quartile=0.03424871, median=, upper quartile=0.0347215, upper whisker=0.0347215, average= },
		] coordinates {};
		\addplot+ [
		draw=olive, fill=lime!40,
		boxplot prepared = { box extend=0.75, draw position=23, lower whisker=180.035, lower quartile=180.035, median=, upper quartile=182.785, upper whisker=182.785, average= },
		] coordinates {};
		\addplot+ [
		draw=olive, fill=lime!40,
		boxplot prepared = { box extend=0.75, draw position=24, lower whisker=0.5978749, lower quartile=0.5978749, median=, upper quartile=0.5983138, upper whisker=0.5983138, average= },
		] coordinates {};
		\addplot+ [
		draw=olive, fill=lime!40,
		boxplot prepared = { box extend=0.75, draw position=25, lower whisker=17.33991, lower quartile=17.33991, median=, upper quartile=204.3685, upper whisker=204.3685, average= },
		] coordinates {};
		\addplot+ [
		draw=olive, fill=lime!40,
		boxplot prepared = { box extend=0.75, draw position=26, lower whisker=0.9367781, lower quartile=0.9367781, median=, upper quartile=0.9780749, upper whisker=0.9780749, average= },
		] coordinates {};
		\addplot+ [
		draw=olive, fill=lime!40,
		boxplot prepared = { box extend=0.75, draw position=27, lower whisker=5.925695, lower quartile=5.925695, median=, upper quartile=6.073396, upper whisker=6.073396, average= },
		] coordinates {};
		\addplot+ [
		draw=olive, fill=lime!40,
		boxplot prepared = { box extend=0.75, draw position=28, lower whisker=0.03424871, lower quartile=0.03424871, median=, upper quartile=0.0347215, upper whisker=0.0347215, average= },
		] coordinates {};
		\addplot+ [
		draw=olive, fill=lime!40,
		boxplot prepared = { box extend=0.75, draw position=29, lower whisker=0.9654641, lower quartile=0.9654641, median=, upper quartile=0.9655018, upper whisker=0.9655018, average= },
		] coordinates {};
		\addplot+ [
		draw=olive, fill=lime!40,
		boxplot prepared = { box extend=0.75, draw position=30, lower whisker=50.32889, lower quartile=50.32889, median=, upper quartile=50.32889, upper whisker=50.32889, average= },
		] coordinates {};
		\addplot+ [
		draw=olive, fill=lime!40,
		boxplot prepared = { box extend=0.75, draw position=31, lower whisker=0.5980929, lower quartile=0.5980929, median=, upper quartile=0.5980929, upper whisker=0.5980929, average= },
		] coordinates {};
		\addplot+ [
		draw=olive, fill=lime!40,
		boxplot prepared = { box extend=0.75, draw position=32, lower whisker=1.003254, lower quartile=1.003254, median=, upper quartile=6.243079, upper whisker=6.243079, average= },
		] coordinates {};
		\addplot+ [
		draw=olive, fill=lime!40,
		boxplot prepared = { box extend=0.75, draw position=33, lower whisker=0.07509538, lower quartile=0.07509538, median=, upper quartile=0.9991958, upper whisker=0.9991958, average= },
		] coordinates {};
		\addplot+ [
		draw=olive, fill=lime!40,
		boxplot prepared = { box extend=0.75, draw position=34, lower whisker=0.07509538, lower quartile=0.07509538, median=, upper quartile=0.9991958, upper whisker=0.9991958, average= },
		] coordinates {};
		\addplot+ [
		draw=olive, fill=lime!40,
		boxplot prepared = { box extend=0.75, draw position=35, lower whisker=49403200.0, lower quartile=49403200.0, median=, upper quartile=104457800.0, upper whisker=104457800.0, average= },
		] coordinates {};
		\addplot+ [
		draw=olive, fill=lime!40,
		boxplot prepared = { box extend=0.75, draw position=36, lower whisker=0.5980929, lower quartile=0.5980929, median=, upper quartile=0.5980929, upper whisker=0.5980929, average= },
		] coordinates {};
		\addplot+ [
		draw=olive, fill=lime!40,
		boxplot prepared = { box extend=0.75, draw position=37, lower whisker=31.8502, lower quartile=31.8502, median=, upper quartile=67.2715, upper whisker=67.2715, average= },
		] coordinates {};
		\addplot+ [
		draw=olive, fill=lime!40,
		boxplot prepared = { box extend=0.75, draw position=38, lower whisker=650943.0, lower quartile=650943.0, median=, upper quartile=650943.0, upper whisker=650943.0, average= },
		] coordinates {};
		\addplot+ [
		draw=olive, fill=lime!40,
		boxplot prepared = { box extend=0.75, draw position=39, lower whisker=1.003254, lower quartile=1.003254, median=, upper quartile=6.243079, upper whisker=6.243079, average= },
		] coordinates {};
		\addplot+ [
		draw=olive, fill=lime!40,
		boxplot prepared = { box extend=0.75, draw position=40, lower whisker=0.9367781, lower quartile=0.9367781, median=, upper quartile=0.9780749, upper whisker=0.9780749, average= },
		] coordinates {};
		\addplot+ [
		draw=olive, fill=lime!40,
		boxplot prepared = { box extend=0.75, draw position=41, lower whisker=1052543000.0, lower quartile=1052543000.0, median=, upper quartile=1054812000.0, upper whisker=1054812000.0, average= },
		] coordinates {};
		\addplot+ [
		draw=olive, fill=lime!40,
		boxplot prepared = { box extend=0.75, draw position=42, lower whisker=24.75, lower quartile=24.75, median=, upper quartile=24.75, upper whisker=24.75, average= },
		] coordinates {};
		\addplot+ [
		draw=olive, fill=lime!40,
		boxplot prepared = { box extend=0.75, draw position=43, lower whisker=5.925695, lower quartile=5.925695, median=, upper quartile=6.073396, upper whisker=6.073396, average= },
		] coordinates {};
		\addplot+ [
		draw=olive, fill=lime!40,
		boxplot prepared = { box extend=0.75, draw position=44, lower whisker=0.03321622, lower quartile=0.03321622, median=, upper quartile=3.004139, upper whisker=3.004139, average= },
		] coordinates {};
		\addplot+ [
		draw=olive, fill=lime!40,
		boxplot prepared = { box extend=0.75, draw position=45, lower whisker=2.543279, lower quartile=2.543279, median=, upper quartile=2.543718, upper whisker=2.543718, average= },
		] coordinates {};
		\addplot+ [
		draw=olive, fill=lime!40,
		boxplot prepared = { box extend=0.75, draw position=46, lower whisker=0.3937781, lower quartile=0.3937781, median=, upper quartile=3.119729, upper whisker=3.119729, average= },
		] coordinates {};
		\addplot+ [
		draw=olive, fill=lime!40,
		boxplot prepared = { box extend=0.75, draw position=47, lower whisker=1052543000.0, lower quartile=1052543000.0, median=, upper quartile=1054812000.0, upper whisker=1054812000.0, average= },
		] coordinates {};
		\addplot+ [
		draw=olive, fill=lime!40,
		boxplot prepared = { box extend=0.75, draw position=48, lower whisker=17.33991, lower quartile=17.33991, median=, upper quartile=204.3685, upper whisker=204.3685, average= },
		] coordinates {};
		\addplot+ [
		draw=olive, fill=lime!40,
		boxplot prepared = { box extend=0.75, draw position=49, lower whisker=1.307142, lower quartile=1.307142, median=, upper quartile=1.307193, upper whisker=1.307193, average= },
		] coordinates {};
		\addplot+ [
		draw=olive, fill=lime!40,
		boxplot prepared = { box extend=0.75, draw position=50, lower whisker=0.8388149, lower quartile=0.8388149, median=, upper quartile=18.83881, upper whisker=18.83881, average= },
		] coordinates {};
	\end{axis}
\end{tikzpicture}

%% file: 5_sat_prec.tex
\begin{tikzpicture}
	\begin{axis}[
		ybar = -7pt,
		width = 1.0\textwidth,
		height = 0.175\textheight,
		symbolic x coords = { 0\_FABS, 1\_COS, 2\_SIN, 3\_FABS, 4\_FMOD, 5\_ATAN2, 6\_ATAN2, 7\_FMOD, 8\_FMOD, 9\_FMOD, 10\_POW, 11\_POW, 12\_FMOD, 13\_FMOD, 14\_FMOD, 15\_ACOS, 16\_SQRT, 17\_FLOOR, 18\_COS, 19\_ACOS, 20\_FABS, 21\_SQRT, 22\_FLOOR, 23\_COS, 24\_COS, 25\_FABS, 26\_SIN, 27\_SQRT, 28\_SQRT, 29\_FLOOR, 30\_COS, 31\_COS, 32\_FABS, 33\_ACOS, 34\_SQRT, 35\_SIN, 36\_SQRT, 37\_SQRT, 38\_SIN, 39\_ACOS, 40\_SQRT, 41\_FLOOR, 42\_COS, 43\_SIN, 44\_FABS, 45\_SIN, 46\_SQRT, 47\_SIN, 48\_SQRT, 49\_FLOOR },
		bar direction = y,
		bar width = 7pt,
		tick align = center,
		enlargelimits = 0.01,
		xtick = data,
		ymin=0,ymax=60,
		ylabel = Vprec-libm output\\ precision (bits),
		y label style = {scale=1.5, transform shape, align=center},
		extra y ticks={23, 28},
		x tick label style = {rotate=45,anchor=east},
		nodes near coords,
		nodes near coords align = {vertical},
		nodes near coords style = {/pgf/number format/set thousands separator={}},
		legend style = {at={(1.14,0.75)}, anchor=east, legend columns=1},
	]
		\addplot[draw=orange, fill=pink!40] coordinates { (0\_FABS, 52) (1\_COS, 52) (2\_SIN, 52) (3\_FABS, 52) (4\_FMOD, 52) (5\_ATAN2, 52) (6\_ATAN2, 52) (7\_FMOD, 52) (8\_FMOD, 52) (9\_FMOD, 52) (10\_POW, 52) (11\_POW, 52) (12\_FMOD, 52) (13\_FMOD, 52) (14\_FMOD, 52) (15\_ACOS, 52) (16\_SQRT, 52) (17\_FLOOR, 52) (18\_COS, 52) (19\_ACOS, 52) (20\_FABS, 52) (21\_SQRT, 52) (22\_FLOOR, 52) (23\_COS, 52) (24\_COS, 52) (25\_FABS, 52) (26\_SIN, 52) (27\_SQRT, 52) (28\_SQRT, 52) (29\_FLOOR, 52) (30\_COS, 52) (31\_COS, 52) (32\_FABS, 52) (33\_ACOS, 52) (34\_SQRT, 52) (35\_SIN, 52) (36\_SQRT, 52) (37\_SQRT, 52) (38\_SIN, 52) (39\_ACOS, 52) (40\_SQRT, 52) (41\_FLOOR, 52) (42\_COS, 52) (43\_SIN, 52) (44\_FABS, 52) (45\_SIN, 52) (46\_SQRT, 52) (47\_SIN, 52) (48\_SQRT, 52) (49\_FLOOR, 52)};
		\addplot[draw=olive, fill=lime!40, nodes near coords style = {yshift=0}] coordinates { (0\_FABS, 0) (1\_COS, 41) (2\_SIN, 42) (3\_FABS, 0) (4\_FMOD, 44) (5\_ATAN2, 29) (6\_ATAN2, 44) (7\_FMOD, 46) (8\_FMOD, 25) (9\_FMOD, 43) (10\_POW, 25) (11\_POW, 43) (12\_FMOD, 44) (13\_FMOD, 0) (14\_FMOD, 43) (15\_ACOS, 23) (16\_SQRT, 28) (17\_FLOOR, 4) (18\_COS, 39) (19\_ACOS, 23) (20\_FABS, 0) (21\_SQRT, 23) (22\_FLOOR, 7) (23\_COS, 42) (24\_COS, 4) (25\_FABS, 0) (26\_SIN, 42) (27\_SQRT, 27) (28\_SQRT, 33) (29\_FLOOR, 4) (30\_COS, 34) (31\_COS, 27) (32\_FABS, 0) (33\_ACOS, 28) (34\_SQRT, 28) (35\_SIN, 24) (36\_SQRT, 35) (37\_SQRT, 44) (38\_SIN, 24) (39\_ACOS, 20) (40\_SQRT, 33) (41\_FLOOR, 1) (42\_COS, 40) (43\_SIN, 23) (44\_FABS, 0) (45\_SIN, 42) (46\_SQRT, 26) (47\_SIN, 11) (48\_SQRT, 39) (49\_FLOOR, 3)};
		\addplot[orange,sharp plot,nodes near coords={},update limits=false,shorten >=-3mm,shorten <=-3mm] coordinates {(0\_FABS,23) (49\_FLOOR,23)};
		\addplot[teal,sharp plot,nodes near coords={},update limits=false,shorten >=-3mm,shorten <=-3mm] coordinates {(0\_FABS,28) (49\_FLOOR,28)};
		\legend{\small Original, \small Optimized}
	\end{axis}
\end{tikzpicture}